\documentstyle{cupconf}
\input{psfig}
\input{epsf}

\ifoldfss
\else
  \ifnfssone
    \newmathalphabet{\mathit}
      \addtoversion{normal}{\mathit}{cmr}{m}{it}
      \addtoversion{bold}{\mathit}{cmr}{bx}{it}
    \newmathalphabet{\mathcal}
      \addtoversion{normal}{\mathcal}{cmsy}{m}{n}
    \else
    \ifnfsstwo
    \fi
  \fi
\fi

\def\apj{{Astrophys. J.}}
\def\apjl{{Astrophys. J. Lett.}}
\def\prl{{Phys. Rev. Lett.}}
\def\prd{{Phys. Rev. D}}

\def\aj{{Astron. J.}}

\def\Mpc{\,{\rm Mpc}}

\def\cmm2{{\,\rm cm^{-2}}}
\def\cm2{{\,{\rm cm}^2}}
\def\cmm3{{\,{\rm cm}^{-3}}}
\def\gcmm3{{\,{\rm g\,cm^{-3}}}}
\def\kms{\,{\rm km\,s^{-1}}}

\def\la{\mathrel{\mathpalette\fun <}}

\def\fun#1#2{\lower3.6pt\vbox{\baselineskip0pt\lineskip.9pt
  \ialign{$\mathsurround=0pt#1\hfil##\hfil$\crcr#2\crcr\sim\crcr}}}

\def\hexnumber#1{\ifcase#1 0\or1\or2\or3\or4\or5\or6\or7\or8\or9\or
 A\or B\or C\or D\or E\or F\fi }

\makeatletter
\ifx\CUP@mtlplain@loaded\undefined
\else
\fi
\makeatother
 \makeatletter
 \ifx\CUP@mtlplain@loaded\undefined
   \font\tenbmi=cmmib10 at 10pt
   \font\sevenbmi=cmmib10 at 7pt
   \font\fivebmi=cmmib10 at 5pt

   \newfam\bmifam
   \textfont\bmifam=\tenbmi
   \scriptfont\bmifam=\sevenbmi
   \scriptscriptfont\bmifam=\fivebmi
   
 \fi
 \makeatother
\ifnfsstwo

\fi
\ifnfssone

\fi
\ifoldfss

\fi

\mathchardef\varLambda="0103

\makeatletter
\ifx\CUP@mtlplain@loaded\undefined
\else
\fi
\makeatother
\makeatletter
\ifx\CUP@mtlplain@loaded\undefined
  \font\tenbms=cmbsy10
  \font\sevenbms=cmbsy10 at 7pt
  \font\fivebms=cmbsy10 at 5pt
  \newfam\bmsfam
  \textfont\bmsfam=\tenbms
  \scriptfont\bmsfam=\sevenbms
  \scriptscriptfont\bmsfam=\fivebms

  \edef\bsy@{\hexnumber\bmsfam}
  \mathchardef\bnabla="0\bsy@72
\fi
\makeatother
\def\etal{\mbox{\it et al.}}

\title[Why Cosmologists Believe the Universe is Accelerating]{Why Cosmologists
Believe the Universe is Accelerating}

\author[M. S. Turner]%
{M\ls I\ls C\ls H\ls A\ls E\ls L\ns S.\ns T\ls U\ls R\ls N\ls E\ls R}

\affiliation{Departments of Astronomy \& Astrophysics and of Physics,
Enrico Fermi Institute\\
The University of Chicago,
Chicago, IL~~60637-1433 USA\\
e-mail:  mturner@oddjob.uchicago.edu\\[\affilskip]
NASA/Fermilab Astrophysics Center, Fermi National Accelerator Laboratory\\
Batavia, IL~~60510-0500 USA}

\begin{document}
\ifnfssone
\else
  \ifnfsstwo
  \else
    \ifoldfss
      \let\mathcal\cal
      \let\mathrm\rm
      \let\mathsf\sf
    \fi
  \fi
\fi

\maketitle

\begin{abstract}

Theoretical cosmologists were quick to be convinced by
the evidence presented in 1998 for the accelerating
Universe.  I explain how this remarkable discovery was the
missing piece in the grand cosmological puzzle.  When
found, it fit perfectly.  For cosmologists, this added
extra weight to the strong evidence of the SN Ia
measurements themselves, making the result all the more
believable.

\end{abstract}

\firstsection % if your document starts with a section,

\section{Introduction}

If theoretical cosmologists are the flyboys of astrophysics,
they were flying on fumes in the 1990s.  Since the early 1980s
inflation and cold dark matter (CDM) have been the dominant theoretical ideas
in cosmology.  However, a key
prediction of inflation, a flat Universe (i.e., $\Omega_0 \equiv
\rho_{\rm total}/\rho_{\rm crit} = 1$), was beginning to look
untenable.  By the late 1990s
it was becoming increasingly clear that matter only accounted for
30\% to 40\% of the critical density (see e.g., Turner, 1999).
Further, the $\Omega_M =1$,
COBE-normalized CDM model was not a very good fit to the data
without some embellishment (15\% or so of the dark matter in neutrinos,
significant deviation from from scale invariance -- called tilt --
or a very low value for the Hubble constant; see e.g., Dodelson
\etal, 1996).

Because of this and their strong belief in inflation,
a number of inflationists (see e.g., Turner, Steigman \& Krauss, 1984
and Peebles, 1984) were led to consider
seriously the possibility that the missing 60\% or so of the critical
density exists in the form of vacuum energy (cosmological
constant) or something even more interesting with similar
properties (see Sec. 3 below).  Since determinations of the matter density
take advantage of its enhanced gravity when it clumps
(in galaxies, clusters or superclusters), vacuum energy, which
is by definition spatially smooth, would not have shown up in the
matter inventory.

Not only did a cosmological constant solve the
``$\Omega$ problem,'' but $\Lambda$CDM, the flat CDM model with
$\Omega_M\sim 0.4$ and $\Omega_\Lambda\sim 0.6$, became the best
fit universe model (Turner, 1991 and 1997b; Krauss \& Turner, 1995;
Ostriker \& Steinhardt, 1995; Liddle \etal, 1996).  In June 1996, at the Critical Dialogues
in Cosmology Meeting at Princeton University,
the only strike recorded against $\Lambda$CDM was the
early SN Ia results of Perlmutter's group (Perlmutter \etal, 1997) which
excluded $\Omega_\Lambda > 0.5$ with 95\% confidence.

The first indirect experimental hint for something like
a cosmological constant came in 1997.  Measurements of the anisotropy of the
cosmic background radiation (CBR) began to show evidence for the
signature of a flat Universe, a peak in the multipole power spectrum at $l=200$.
Unless the estimates of the matter density were wildly wrong,
this was evidence for a smooth, dark energy
component.  A universe with $\Omega_\Lambda
\sim 0.6$ has a smoking gun signature:  it is speeding up
rather than slowing down.  In 1998 came the SN Ia evidence
that our Universe is speeding up; for some cosmologists this was
a great surprise.  For many theoretical cosmologists
this was the missing piece of the grand puzzle
and the confirmation of a prediction.

\section{The theoretical case for accelerated expansion}

The case for accelerated expansion that existed in January
1998 had three legs:  growing evidence that $\Omega_M
\sim 0.4$ and not 1; the inflationary prediction of a flat
Universe and hints from CBR anisotropy that this was indeed
true; and the failure of simple $\Omega_M =1$ CDM model
and the success of $\Lambda$CDM.  The tension between measurements
of the Hubble constant and age determinations for the oldest
stars was also suggestive, though because of the uncertainties,
not as compelling.  Taken together, they foreshadowed the
presence of a cosmological constant (or something similar) and the
discovery of accelerated expansion.

To be more precise, Sandage's deceleration parameter is given by
\begin{equation}
q_0 \equiv {(\ddot R /R)_0 \over H_0^2}
        = {1 \over 2}\Omega_0 + {3\over 2} \sum_i \Omega_i w_i \,,
\end{equation}
where the pressure of component $i$, $p_i \equiv w_i \rho_i$;
e.g., for baryons $w_i = 0$, for radiation $w_i = 1/3$,
and for vacuum energy $w_X = -1$.  For $\Omega_0 = 1$,
$\Omega_M =0.4$ and $w_X < -{5\over 9}$, the deceleration
parameter is negative.  The kind of dark component needed to
pull cosmology together implies accelerated expansion.

\subsection{Matter/energy inventory:  $\Omega_0 =1\pm 0.2$,
$\Omega_M=0.4\pm 0.1$}

There is a growing consensus that the anisotropy of the CBR
offers the best means of determining the curvature of the Universe
and thereby $\Omega_0$.
This is because the method is intrinsically geometric --
a standard ruler on the last-scattering surface --
and involves straightforward physics at
a simpler time (see e.g., Kamionkowski \etal, 1994).
It works like this.

\begin{figure}
\centerline{\psfig{figure=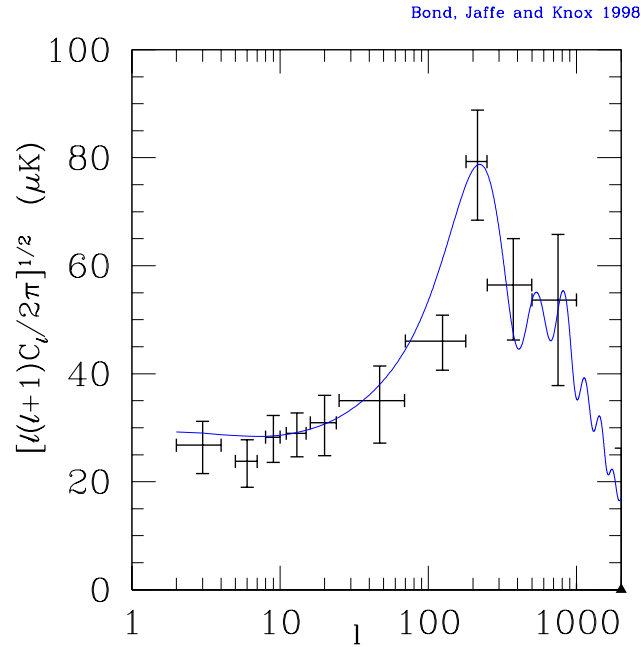,width=3.5in}}
\caption{Current CBR anisotropy data, averaged and binned to
reduce error bars and visual confusion.  The theoretical
curve is for the $\Lambda$CDM model with $H_0=65\,{\rm km\,
s^{-1}\,Mpc^{-1}}$ and $\Omega_M =0.4$; note the
goodness of fit (Figure courtesy of L. Knox).
}
\label{fig:cbr_knox}
\end{figure}

At last scattering baryonic matter (ions and electrons) was still tightly
coupled to photons; as the baryons fell into the dark-matter
potential wells the pressure of photons acted as a restoring
force, and gravity-driven acoustic oscillations resulted.  These
oscillations can be decomposed into their Fourier modes;
Fourier modes with $k\sim l H_0/2$ determine the multipole
amplitudes $a_{lm}$ of CBR anisotropy.  Last scattering
occurs over a short time, making the CBR is a snapshot of
the Universe at $t_{\rm ls} \sim 300,000\,$yrs.
Each mode is ``seen'' in a well defined phase of its
oscillation.  (For the density perturbations predicted by inflation,
all modes the have same initial phase because
all are growing-mode perturbations.)
Modes caught at maximum compression or rarefaction lead
to the largest temperature anisotropy; this results in a series of
acoustic peaks beginning at $l\sim 200$ (see Fig.~\ref{fig:cbr_knox}).
The  wavelength of the lowest
frequency acoustic mode that has reached maximum compression,
$\lambda_{\rm max} \sim v_s t_{\rm ls}$, is the standard
ruler on the last-scattering surface.  Both $\lambda_{\rm
max}$ and the distance to the last-scattering surface depend
upon $\Omega_0$, and the position of the first peak $l\simeq
200/\sqrt{\Omega_0}$.  This relationship is insensitive
to the composition of matter and energy in the Universe.

CBR anisotropy measurements, shown in Fig.~\ref{fig:cbr_knox},
now cover three orders of magnitude in multipole and are from
more than twenty experiments.  COBE is
the most precise and covers multipoles $l=2-20$;
the other measurements come from balloon-borne, Antarctica-based
and ground-based experiments using both low-frequency
($f<100\,$GHz) HEMT receivers and high-frequency ($f>100\,$GHz)
bolometers.  Taken together, all the measurements are beginning to
define the position of the first acoustic peak, at a value that is
consistent with a flat Universe.  Various analyses of the
extant data have been carried out, indicating $\Omega_0 \sim 1\pm 0.2$
(see e.g., Lineweaver, 1998).
It is certainly too early to draw definite conclusions or put
too much weigh in the error estimate.  However, a strong
case is developing for a flat Universe and more data is on
the way (Python V, Viper, MAT, Maxima, Boomerang, CBI, DASI, and others).
Ultimately, the issue will be settled by NASA's
MAP (launch late 2000) and ESA's Planck (launch 2007) satellites
which will map the entire CBR sky with 30 times the resolution
of COBE (around $0.1^\circ$) (see Page and Wilkinson, 1999).

Since the pioneering work of Fritz Zwicky and Vera Rubin, it has been
known that there is far too little material in the form of stars
(and related material)
to hold galaxies and clusters together, and thus, that
most of the matter in the Universe is dark (see e.g. Trimble, 1987).
Weighing the dark matter has been the challenge.
At present, I believe that clusters provide the most reliable
means of estimating the total matter density.

Rich clusters are relatively rare objects -- only
about 1 in 10 galaxies is found in a rich cluster --
which formed from density perturbations of (comoving) size
around 10\,Mpc.  However, because they gather together material
from such a large region of space, they can provide
a ``fair sample'' of matter in the Universe.  Using clusters as such,
the precise BBN baryon density can be used
to infer the total matter density (White \etal, 1993).
(Baryons and dark matter need not be
well mixed for this method to work
provided that the baryonic and total mass are determined
over a large enough portion of the cluster.)

Most of the baryons in clusters reside in the
hot, x-ray emitting intracluster gas and not in the galaxies
themselves, and so the problem essentially reduces to determining
the gas-to-total mass ratio.
The gas mass can be determined by two methods:
1) measuring the x-ray flux from the intracluster
gas and 2) mapping the Sunyaev - Zel'dovich
CBR distortion caused by CBR photons scattering off hot electrons in the
intracluster gas.  The total cluster mass can be determined
three independent ways:  1)  using the motions of clusters galaxies
and the virial theorem; 2) assuming that the gas is in hydrostatic
equilibrium and using it to infer the underlying mass distribution; and
3) mapping the cluster mass directly by gravitational lensing
(Tyson, 1999).  Within their
uncertainties, and where comparisons can be made, the three methods
for determining the total mass agree (see e.g., Tyson, 1999);
likewise, the two methods for determining the gas mass are consistent.

\begin{figure}
\centerline{\psfig{figure=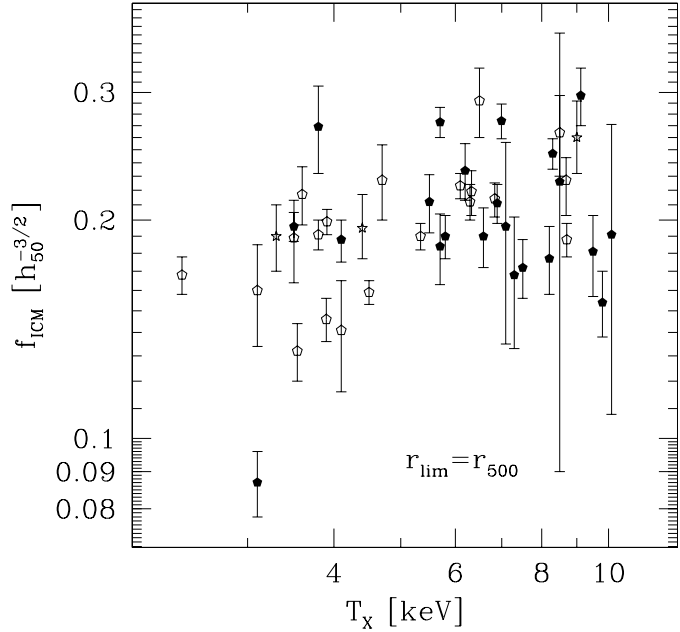,width=3.5in}}
\caption{Cluster gas fraction as a function of
cluster gas temperature for a sample of 45 galaxy clusters
(Mohr \etal, 1998).  While there is some indication
that the gas fraction
decreases with temperature for $T< 5\,$keV, perhaps because
these lower-mass clusters lose some of their hot gas, the
data indicate that the gas fraction reaches a plateau
at high temperatures, $f_{\rm gas} =0.212 \pm 0.006$
for $h=0.5$ (Figure courtesy of Joe Mohr).
}
\label{fig:gas}
\end{figure}

Mohr \etal\ (1998) have compiled the gas to total mass ratios determined from
x-ray measurements for a sample of 45 clusters; they find
$f_{\rm gas} = (0.075\pm 0.002)h^{-3/2}$ (see Fig.~\ref{fig:gas}).
Carlstrom (1999), using his
S-Z gas measurements and x-ray measurements for the total
mass for 27 clusters, finds $f_{\rm gas} =(0.06\pm 0.006)h^{-1}$.
(The agreement of these two numbers means that clumping of
the gas, which could lead to an overestimate of the gas fraction
based upon the x-ray flux, is not a problem.)
Invoking the ``fair-sample assumption,'' the mean
matter density in the Universe can be inferred:
\begin{eqnarray}
\Omega_M = \Omega_B/f_{\rm gas} & = & (0.3\pm 0.05)h^{-1/2}\  ({\rm X ray})\nonumber\\
                   & = &  (0.25\pm 0.04)h^{-1}\ ({\rm S-Z}) \nonumber \\
                   & = & 0.4\pm 0.1\ ({\rm my\ summary})\,.
\end{eqnarray}

I believe this to be the most reliable and precise
determination of the matter density.  It involves few assumptions,
most of which have now been tested.  For example, the agreement
of S-Z and x-ray gas masses implies that gas clumping is not
significant; the agreement of x-ray and lensing estimates for
the total mass implies that hydrostatic equilibrium is a good
assumption; the gas fraction does not vary significantly
with cluster mass.

%%\begin{figure}
%%\centerline{\psfig{figure=omega_sum.eps,width=4.5in}}
%%\caption{Summary of matter/energy in the Universe.
%%The right side refers to an overall accounting of matter
%%and energy; the left refers to the composition of the matter
%%component.  The contribution of relativistic particles,
%%CBR photons and neutrinos, $\Omega_{\rm rel}h^2 = 4.170\times
%%10^{-5}$, is not shown.  The upper limit to mass density contributed
%%by neutrinos is based upon the failure of the hot dark
%%matter model of structure formation (White, Frenk \& Davis,
%%1983; and Dodelson \etal, 1996) and the lower limit follows from the
%%evidence for neutrino oscillations (Fukuda \etal, 1998).
%%Here $H_0$ is taken to be $65\,{\rm km\,s^{-1}\,Mpc^{-1}}$.
%%}
%%\label{fig:omega}
%%\end{figure}

\subsection{Dark energy}

The apparently contradictory results, $\Omega_0 = 1\pm 0.2$
and $\Omega_M = 0.4\pm 0.1$,
can be reconciled by the presence of a dark-energy component
that is nearly smoothly distributed.  The cosmological
constant is the simplest possibility and it has $p_X=-\rho_X$.
There are other possibilities for the smooth, dark energy.
As I now discuss, other constraints imply that such a component
must have very negative pressure ($w_X \la -{1\over 2}$)
leading to the prediction of accelerated expansion.

\begin{figure}
\centerline{\psfig{figure=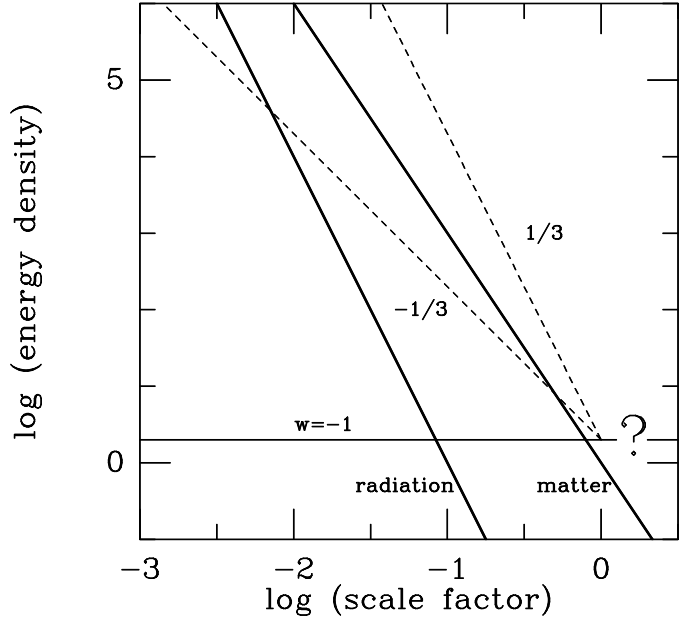,width=3.5in}}
\caption{Evolution of the energy density in matter,
radiation (heavy lines), and different possibilities
for the dark-energy component ($w=-1,-{1\over 3},{1\over 3}$)
vs. scale factor.  The matter-dominated era begins
when the scale factor was $\sim 10^{-4}$ of its present
size (off the figure) and ends when the dark-energy
component begins to dominate, which depends upon the
value of $w$:  the more negative $w$ is, the longer
the matter-dominated era in which density perturbations
can go into the large-scale structure seen today.
These considerations require $w<-{1\over 3}$ (Turner
\& White, 1997).
}
\label{fig:xmatter}
\end{figure}

To begin, parameterize the bulk equation of state of
this unknown component:  $w \equiv p_X/\rho_X$ (Turner \& White,
1997).  This implies
that its energy density evolves as $\rho_X \propto
R^{-n}$ where $n=3(1+w)$.  The development
of the structure observed today from density perturbations of the
size inferred from measurements of the anisotropy of the CBR
requires that the Universe be matter dominated from the epoch
of matter -- radiation equality until very recently.  Thus,
to avoid interfering with structure formation, the dark-energy component
must be less important in the past than it is today.
This implies that $n$ must be less than $3$ or $w< 0$; the more negative
$w$ is, the faster this component gets out of the way (see
Fig.~\ref{fig:xmatter}).  More careful consideration of the
growth of structure implies that $w$ must be less than about
$-{1\over 3}$ (Turner \& White, 1997).

\begin{figure}
\centerline{\psfig{figure=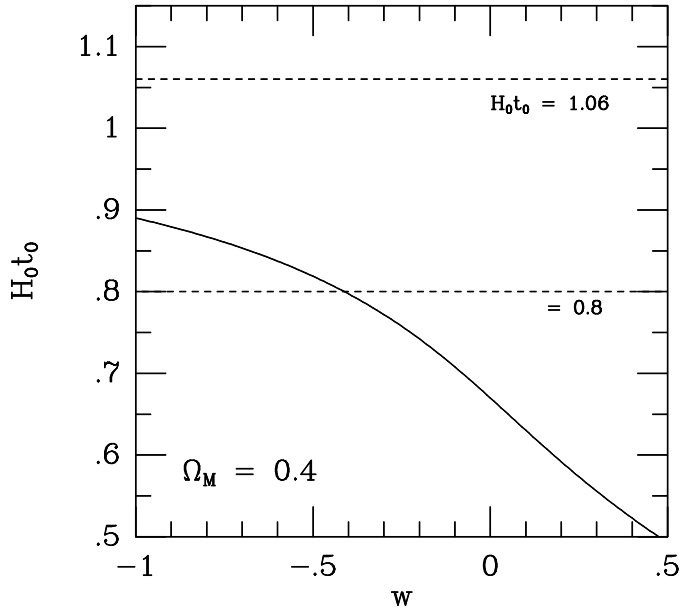,width=3.5in}}
\caption{$H_0t_0$ vs. the equation of state for the
dark-energy component.  As can be seen, an added benefit of
a component with negative pressure is an older Universe for a given
Hubble constant.  The broken horizontal lines denote the
$1\sigma$ range for $H_0=65\pm 5\,{\rm km\,s^{-1}\,Mpc^{-1}}$
and $t_0=14\pm 1.5\,$Gyr, and indicate that $w<-{1\over 2}$ is
preferred.
}
\label{fig:wage}
\end{figure}

Next, consider the constraint provided by the age of the Universe
and the Hubble constant.  Their product, $H_0t_0$, depends the
equation of state of the Universe; in particular, $H_0t_0$ increases with
decreasing $w$ (see Fig.~\ref{fig:wage}).  To be definite, I will take $t_0
=14\pm 1.5\,$Gyr and $H_0=65\pm 5\,{\rm km\,s^{-1}\,Mpc^{-1}}$
(see e.g., Chaboyer \etal, 1998 and Freedman, 1999); this implies
that $H_0t_0 = 0.93 \pm 0.13$.  Fig.~\ref{fig:wage} shows that
$w<-{1\over 2}$ is preferred by age/Hubble constant considerations.

To summarize, consistency between $\Omega_M \sim 0.4$ and
$\Omega_0 \sim 1$ along with other cosmological considerations
implies the existence of a dark-energy component with bulk
pressure more negative than about $-\rho_X /2$.  The simplest
example of such is vacuum energy (Einstein's cosmological
constant), for which $w=-1$.  The smoking-gun signature of
a smooth, dark-energy component is accelerated expansion since
$q_0 = 0.5 + 1.5w_X\Omega_X \simeq 0.5 + 0.9w < 0$ for
$w<-{5\over 9}$.

\subsection{$\Lambda$CDM}

The cold dark matter scenario for structure formation is
the most quantitative and most successful model ever proposed.
Two of its key features are inspired by inflation:  almost scale invariant, adiabatic
density perturbations with Gaussian statistical properties and
a critical density Universe.  The third, nonbaryonic dark matter
is a logical consequence of the inflationary prediction of a flat
universe and the BBN-determination of the baryon density at 5\%
of the critical density.

There is a very large body of data that is consistent with it:
the formation epoch of galaxies and distribution of galaxy
masses, galaxy correlation function and its evolution,
abundance of clusters and its evolution, large-scale structure,
and on and on.  In the early 1980s attention was focused on a
``standard CDM model'':  $\Omega_0=\Omega_M =1$, $\Omega_B =
0.05$, $h=0.50$, and exactly scale invariant density perturbations
(the cosmological equivalent of DOS 1.0).  The detection of CBR anisotropy
by COBE DMR in 1992 changed everything.

First and most importantly, the COBE DMR detection
validated the gravitational instability
picture for the growth of large-scale structure:  The level of
matter inhomogeneity implied at last scattering, after 14 billion
years of gravitational amplification, was consistent with the structure
seen in the Universe today.  Second, the anisotropy, which was detected
on the $10^\circ$ angular scale, permitted an accurate normalization
of the CDM power spectrum.  For ``standard cold dark matter'', this
meant that the level of inhomogeneity on all scales could be accurately
predicted.  It turned out to be about a factor of two too large on
galactic scales.  Not bad for an ab initio theory.

With the COBE detection came the realization that the quantity and
quality of data that bear on CDM was increasing and that the theoretical
predictions would have to match their precision.  Almost overnight,
CDM became a ten (or so) parameter theory.
For astrophysicists, and especially cosmologists,
this is daunting, as it may seem that a ten-parameter
theory can be made to fit any set of observations.  This is
not the case when one has the quality and quantity of data
that will soon be available.

In fact, the ten parameters of CDM + Inflation
are an opportunity rather than a curse:  Because the parameters
depend upon the underlying inflationary model and fundamental
aspects of the Universe, we have the very real possibility of learning
much about the Universe and inflation.  The ten parameters
can be organized into two groups:  cosmological
and dark-matter (Dodelson \etal, 1996).

\smallskip
\centerline{\it Cosmological Parameters}
\vspace{3pt}
\begin{enumerate}

\item $h$, the Hubble constant in units of $100\kms\Mpc^{-1}$.

\item $\Omega_Bh^2$, the baryon density.  Primeval deuterium
measurements and together with the theory of BBN imply:
$\Omega_Bh^2 = 0.02 \pm 0.002$.

\item $n$, the power-law index of the scalar density perturbations.
CBR measurements indicate $n=1.1\pm 0.2$; $n=1$ corresponds to
scale-invariant density perturbations.  Many
inflationary models predict $n\simeq 0.95$; range of predictions
runs from $0.7$ to $1.2$.

\item $dn/d\ln k$, ``running'' of the scalar index with comoving scale
($k=$ wavenumber).  Inflationary models predict a value of
${\cal O}(\pm 10^{-3})$ or smaller.

\item $S$, the overall amplitude squared of density perturbations,
quantified by their contribution to the variance of the
CBR quadrupole anisotropy.

\item $T$, the overall amplitude squared of gravity waves,
quantified by their contribution to the variance of the
CBR quadrupole anisotropy.  Note, the COBE normalization determines
$T+S$ (see below).

\item $n_T$, the power-law index of the gravity wave spectrum.
Scale-invariance corresponds to $n_T=0$; for inflation, $n_T$
is given by $-{1\over 7}{T\over S}$.

\end{enumerate}

\smallskip
\centerline{\it Dark-matter Parameters}
\vspace{3pt}

\begin{enumerate}

\item $\Omega_\nu$, the fraction of critical density in neutrinos
($=\sum_i m_{\nu_i}/90h^2$).  While the hot dark matter theory of structure
formation is not viable, we now know that neutrinos contribute
at least 0.3\% of the critical density (Fukuda \etal, 1998).

\item $\Omega_X$ and $w_X$, the fraction of critical density in
a smooth dark-energy component and its equation of state.
The simplest example is a cosmological constant ($w_X = -1$).

\item $g_*$, the quantity that counts the number of ultra-relativistic
degrees of freedom.  The standard cosmology/standard
model of particle physics predicts $g_* = 3.3626$.
The amount of radiation controls when
the Universe became matter dominated and thus affects the present
spectrum of density inhomogeneity.

\end{enumerate}

\begin{figure}
\centerline{\psfig{figure=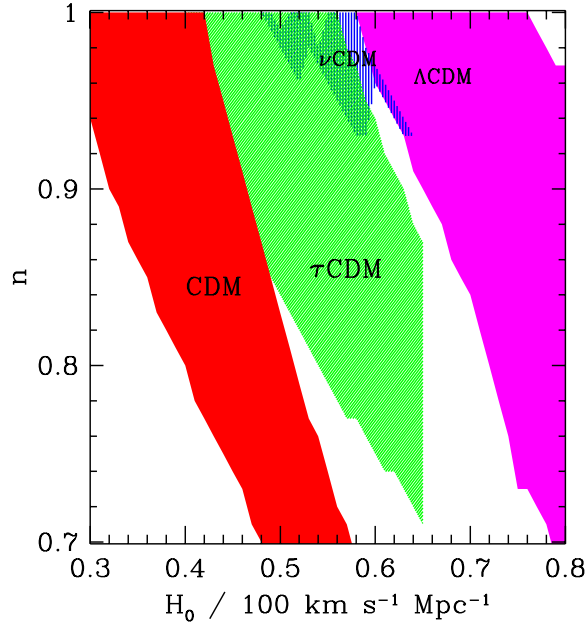,width=3in}}
\caption{Summary of viable CDM models, based upon
CBR anisotropy and determinations of the present
power spectrum of inhomogeneity (Dodelson \etal, 1996).}
\label{fig:cdm_sum}
\end{figure}

A useful way to organize the different CDM models is by their
dark-matter content; within each CDM family, the cosmological
parameters vary.  One list of models is:

\begin{enumerate}

\item sCDM (for simple):  Only CDM and baryons; no additional
radiation ($g_*=3.36$).  The original standard CDM is a member
of this family ($h=0.50$, $n=1.00$, $\Omega_B=0.05$), but is
now ruled out (see Fig.~\ref{fig:cdm_sum}).

\item $\tau$CDM:  This model has
extra radiation, e.g., produced by the decay of an unstable
massive tau neutrino (hence the name); here we take $g_* = 7.45$.

\item $\nu$CDM (for neutrinos):  This model has a dash of hot
dark matter; here we take $\Omega_\nu = 0.2$ (about 5\,eV
worth of neutrinos).

\item $\Lambda$CDM (for cosmological constant) or more generally
xCDM:  This model has a smooth dark-energy component; here,
we take $\Omega_X = \Omega_\Lambda = 0.6$.

\end{enumerate}

Figure \ref{fig:cdm_sum} summarizes the viability of these different CDM models,
based upon CBR measurements and current determinations of
the present power spectrum of inhomogeneity (derived from
redshift surveys).   sCDM is only viable for low values of the
Hubble constant (less than $55\kms\Mpc^{-1}$) and/or
significant tilt (deviation from scale invariance); the region
of viability for $\tau$CDM is similar to sCDM, but shifted
to larger values of the Hubble constant (as large as
$65\kms\Mpc^{-1}$).  $\nu$CDM has an island of viability
around $H_0\sim 60\kms\Mpc^{-1}$ and $n\sim 0.95$.  $\Lambda$CDM
can tolerate the largest values of the Hubble constant.
While the COBE DMR detection ruled out ``standard CDM,''
a host of attractive variants were still viable.

However, when other very relevant data are considered too -- e.g.,
age of the Universe, determinations of the cluster baryon fraction,
measurements of the Hubble constant, and limits to
$\Omega_\Lambda$ -- $\Lambda$CDM emerges as the hands-down-winner
of ``best-fit CDM model'' (Krauss \& Turner, 1995;
Ostriker \& Steinhardt, 1995; Liddle \etal, 1996;
Turner, 1997b).  At the time of the Critical Dialogues in Cosmology
meeting in 1996, the only strike against $\Lambda$CDM was the
absence of evidence for its smoking gun signature, accelerated
expansion.

\begin{figure}
\centerline{\psfig{figure=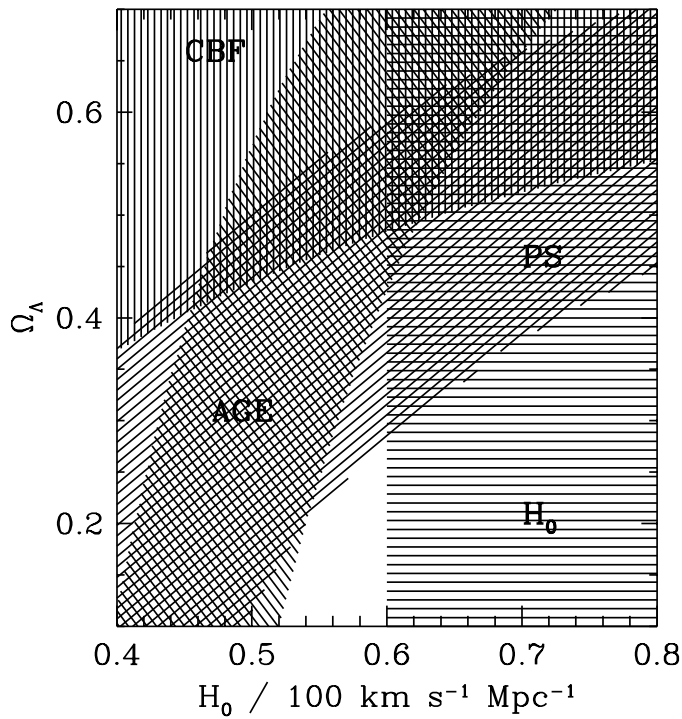,width=3.5in}}
\caption{Constraints used to determine the best-fit CDM model:
PS = large-scale structure + CBR anisotropy; AGE = age of the
Universe; CBF = cluster-baryon fraction; and $H_0$= Hubble
constant measurements.  The best-fit model, indicated by
the darkest region, has $H_0\simeq 60-65\,{\rm km\,s^{-1}
\,Mpc^{-1}}$ and $\Omega_\Lambda
\simeq 0.55 - 0.65$.  Evidence for its smoking-gun signature --
accelerated expansion -- was presented in 1998 (adapted
from Krauss \& Turner, 1995 and Turner, 1997).}
\label{fig:best_fit}
\end{figure}

\subsection{Missing energy found!}

In 1998 evidence for the accelerated expansion anticipated
by theorists was presented in the form
of the magnitude -- redshift (Hubble)
diagram for fifty-some type Ia supernovae (SNe Ia)
out to redshifts of nearly 1.
Two groups, the Supernova Cosmology Project (Perlmutter \etal, 1998)
and the High-z Supernova Search Team (Riess \etal, 1998),
working independently and using different
methods of analysis, each found evidence for accelerated expansion.
Perlmutter \etal\ (1998) summarize their results as a constraint
to a cosmological constant (see Fig.~\ref{fig:omegalambda}),
\begin{equation}
\Omega_\Lambda = {4\over 3}\Omega_M +{1\over 3} \pm {1\over 6}\,.
\end{equation}
For $\Omega_M\sim 0.4 \pm 0.1$, this implies $\Omega_\Lambda =
0.85 \pm 0.2$, or just what is needed to account for the missing energy!
As I have tried to explain, cosmologists were quick than
most to believe, as accelerated expansion was the missing piece
of the puzzle found.

Recently, two other studies, one based upon the x-ray properties of
rich clusters of galaxies (Mohr \etal, 1999) and the other based
upon the properties of double-lobe radio galaxies
(Guerra \etal, 1998), have reported evidence
for a cosmological constant (or similar dark-energy component)
that is consistent with the SN Ia results (i.e., $\Omega_\Lambda \sim 0.7$).

There is another test of an accelerating Universe whose
results are more ambiguous.  It is based upon the fact that
the frequency of multiply lensed
QSOs is expected to be significantly higher in an accelerating
universe (Turner, 1990).  Kochanek (1996) has used gravitational
lensing of QSOs to place a
95\% cl upper limit, $\Omega_\Lambda < 0.66$; and Waga and Miceli
(1998) have generalized it to a dark-energy component with negative
pressure:  $\Omega_X < 1.3 + 0.55w$ (95\% cl), both results for a flat
Universe.  On the other hand, Chiba and Yoshii (1998) claim evidence
for a cosmological constant, $\Omega_\Lambda = 0.7^{+0.1}_{-0.2}$,
based upon the same data.  From this I conclude:  1) Lensing
excludes $\Omega_\Lambda$ larger than 0.8; 2) Because of the
modeling uncertainties and lack of sensitivity
for $\Omega_\Lambda <0.55$, lensing has little power in
strictly constraining $\Lambda$ or a dark component; and 3) When
larger objective surveys of gravitational-lensed
QSOs are carried out (e.g., the Sloan Digital Sky Survey),
there is the possibility of uncovering another smoking-gun
for accelerated expansion.

\subsection{Cosmic concordance}

\begin{figure}
\centerline{\psfig{figure=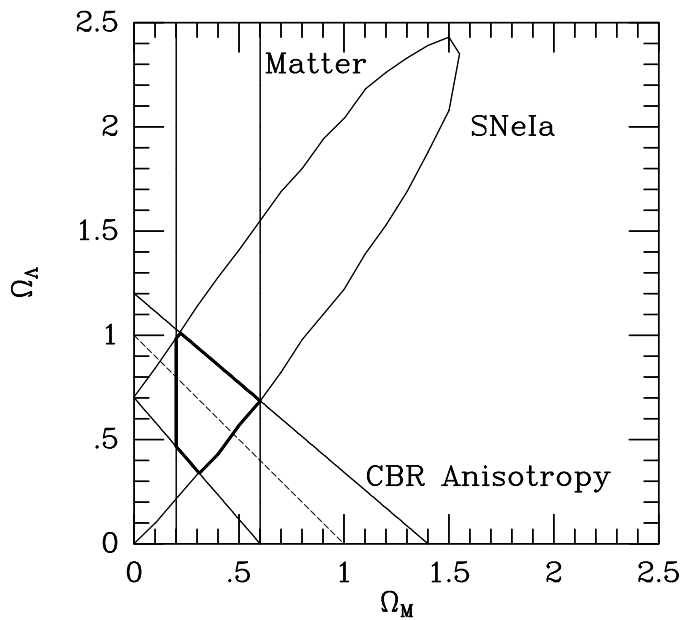,width=3.5in}}
\caption{Two-$\sigma$ constraints to $\Omega_M$ and $\Omega_\Lambda$
from CBR anisotropy, SNe Ia, and measurements of clustered matter.
Lines of constant $\Omega_0$ are diagonal, with a flat
Universe shown by the broken line.
The concordance region is shown in bold:  $\Omega_M\sim 1/3$,
$\Omega_\Lambda \sim 2/3$, and $\Omega_0 \sim 1$.
(Particle physicists who rotate the figure by $90^\circ$
will recognize the similarity to the convergence of the
gauge coupling constants.)
}
\label{fig:omegalambda}
\end{figure}

With the SN Ia results we have for the first time a complete
and self-consistent
accounting of mass and energy in the Universe.
%% (see Fig.~\ref{fig:omega}).
The consistency of the matter/energy accounting
is illustrated in Fig.~\ref{fig:omegalambda}.
Let me explain this exciting figure.  The SN Ia results are sensitive to the
acceleration (or deceleration) of the expansion
and constrain the combination ${4\over 3}\Omega_M -\Omega_\Lambda$.  (Note,
$q_0 = {1\over 2}\Omega_M - \Omega_\Lambda$; ${4\over 3}\Omega_M -
\Omega_\Lambda$ corresponds to the deceleration parameter
at redshift $z\sim 0.4$, the median redshift of these
samples).  The (approximately) orthogonal combination,
$\Omega_0 = \Omega_M + \Omega_\Lambda$
is constrained by CBR anisotropy.  Together, they define a concordance
region around $\Omega_0\sim 1$, $\Omega_M \sim 1/3$,
and $\Omega_\Lambda \sim 2/3$.  The constraint
to the matter density alone, $\Omega_M = 0.4\pm 0.1$,
provides a cross check, and it is consistent with these numbers.
Further, these numbers point to $\Lambda$CDM (or something similar)
as the cold dark matter model.  Another body of observations
already support this as the best fit model.  Cosmic concordance
indeed!

\section{What is the dark energy?}

I have often used the term exotic to refer to particle
dark matter.  That term will now have to be reserved for the
dark energy that is causing the accelerated expansion of the
Universe -- by any standard, it is more exotic and more
poorly understood.  Here is what we do know:   it contributes
about 60\% of the critical density; it has pressure more negative
than about $-\rho /2$; and it does not clump
(otherwise it would have contributed to estimates
of the mass density).  The simplest possibility is
the energy associated with the virtual particles that populate
the quantum vacuum; in this case $p=-\rho$ and the dark energy
is absolutely spatially and temporally uniform.

This ``simple'' interpretation has its difficulties.   Einstein ``invented''
the cosmological constant to make a static model of the Universe
and then he discarded it; we now know that the concept is not optional.
The cosmological constant corresponds to the energy associated
with the vacuum.  However, there is no sensible calculation of
that energy (see e.g., Zel'dovich, 1967; Bludman and Ruderman,
1977; and Weinberg, 1989),
with estimates ranging from $10^{122}$ to $10^{55}$ times the critical
density.  Some particle physicists believe that when the
problem is understood, the answer will be zero.  Spurred
in part by the possibility that cosmologists may have actually weighed
the vacuum (!), particle theorists are taking a fresh look
at the problem (see e.g., Harvey, 1998; Sundrum, 1997).  Sundrum's
proposal, that the gravitational energy of the vacuum is close to the present
critical density because the graviton is a composite particle
with size of order 1\,cm, is indicative of the profound
consequences that a cosmological constant has for fundamental physics.

Because of the theoretical problems mentioned above, as well as the
checkered history of the cosmological constant, theorists have
explored other possibilities for a smooth, component to the dark
energy (see e.g., Turner \& White, 1997).
Wilczek and I pointed out that even if the
energy of the true vacuum is zero, as the Universe as
cooled and went through a series of phase transitions, it
could have become hung up in a metastable vacuum with
nonzero vacuum energy (Turner \& Wilczek, 1982).  In the
context of string theory, where there are a very large number
of energy-equivalent vacua, this becomes a more
interesting possibility:  perhaps the degeneracy of vacuum
states is broken by very small effects, so small that
we were not steered into the lowest energy vacuum during
the earliest moments.

Vilenkin (1984) has suggested a tangled network of very light
cosmic strings (also see, Spergel \& Pen, 1997) produced
at the electroweak phase transition; networks of other frustrated
defects (e.g., walls) are also possible.  In general, the
bulk equation-of-state of frustrated defects
is characterized by $w=-N/3$ where $N$
is the dimension of the defect ($N=1$ for strings, $=2$
for walls, etc.).  The SN Ia data almost exclude strings,
but still allow walls.

An alternative that has received a lot of attention is
the idea of a ``decaying cosmological constant'', a termed
coined by the Soviet cosmologist Matvei Petrovich Bronstein in 1933
(Bronstein, 1933).  (Bronstein was executed on Stalin's
orders in 1938, presumably for reasons not directly related to the
cosmological constant; see Kragh, 1996.)
The term is, of course, an oxymoron; what people have in mind
is making vacuum energy dynamical.  The simplest realization
is a dynamical, evolving scalar field.  If it is spatially homogeneous,
then its energy density and pressure are given by
\begin{eqnarray}
\rho & = & {1\over 2}{\dot\phi}^2 + V(\phi ) \nonumber \\
p    & = & {1\over 2}{\dot\phi}^2 - V(\phi )
\end{eqnarray}
and its equation of motion by (see e.g., Turner, 1983)
\begin{equation}
\ddot \phi + 3H\dot\phi + V^\prime (\phi ) = 0
\end{equation}

The basic idea is that energy of the true vacuum is zero, but
not all fields have evolved to their state of minimum
energy.  This is qualitatively different from that
of a metastable vacuum, which is a local minimum of the
potential and is classically stable.  Here, the field is
classically unstable and is rolling toward its lowest
energy state.

\begin{figure}
\centerline{\epsfxsize=12cm \epsfbox{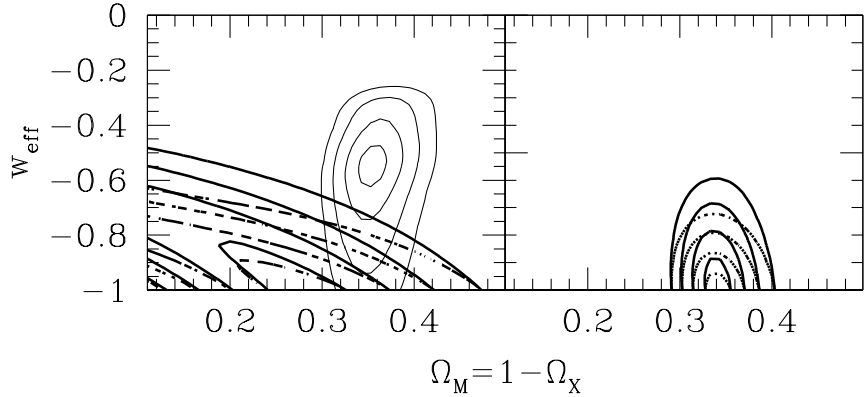}}
\caption{Contours of likelihood, from $0.5\sigma$ to $2\sigma$, in the
$\Omega_M$--$w_{\rm eff}$ plane.  Left:  The thin solid lines are the
constraints from LSS and the CMB.  The heavy lines are the SN Ia
constraints for constant $w$ models (solid curves)
and for a scalar-field model
with an exponential potential (broken curves).
Right:  The likelihood contours from all of our
cosmological constraints for constant $w$ models (solid)
and dynamical scalar-field models (broken).  Note:  at 95\% cl
$w_{\rm eff}$ must be less $-0.6$, and
the cosmological constant is the most likely solution
(from Perlmutter, Turner \& White, 1999).}
\label{fig:composite}
\end{figure}

Two features of the ``rolling-scalar-field scenario'' are
worth noting.  First, the effective equation of state,
$w=({1\over 2}\dot\phi^2 - V)/({1\over 2}\dot\phi^2 +V)$,
can take on any value from 1 to $-1$.  Second, $w$ can
vary with time.  These are key features that may allow it
to be distinguished from the other possibilities.  The
combination of SN Ia, CBR and large-scale structure data
are already beginning to significantly constrain models
(Perlmutter, Turner \& White, 1999), and interestingly enough,
the cosmological constant is still the best fit
(see Fig.~\ref{fig:composite}).

The rolling scalar field scenario (aka mini-inflation
or quintessence) has received a lot of attention over
the past decade (Freese \etal, 1987; Ozer \& Taha, 1987;
Ratra \& Peebles, 1988; Frieman \etal, 1995; Coble \etal, 1996;
Turner \& White, 1997; Caldwell \etal, 1998; Steinhardt, 1999).
It is an interesting idea,
but not without its own difficulties.  First,
one must {\em assume} that the energy of the
true vacuum state ($\phi$ at the minimum of its potential)
is zero; i.e., it does not address the cosmological
constant problem.  Second, as Carroll (1998) has emphasized,
the scalar field is very light and can mediate long-range forces.
This places severe constraints on it.  Finally,
with the possible exception of one model (Frieman \etal,
1995), none of the
scalar-field models address how $\phi$ fits into the
grander scheme of things and why it is so light ($m\sim 10^{-33}\,$eV).

\section{Looking ahead}

Theorists often require new results to pass Eddington's test:
No experimental result should be believed until confirmed by theory.
While provocative (as Eddington had apparently intended it
to be), it embodies the wisdom of mature science.  Results that
bring down the entire conceptual framework are very rare indeed.

\begin{figure}
\centerline{\psfig{figure=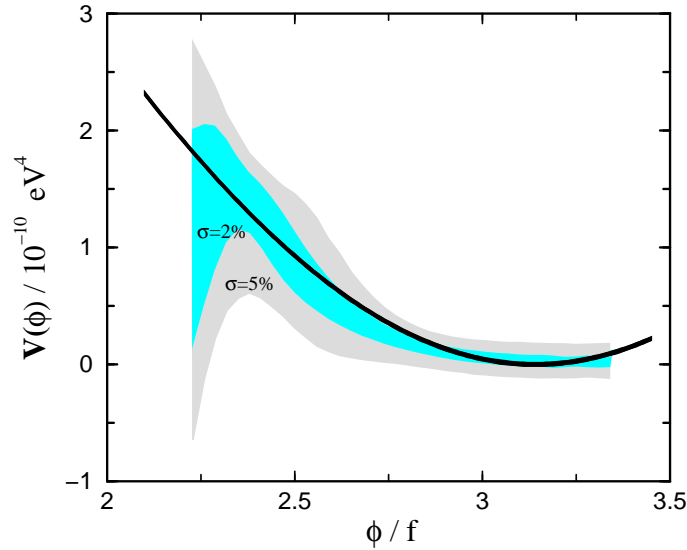,width=3.5in}}
\caption{The 95\% confidence interval for the reconstructed
potential assuming luminosity distance errors of 5\% and 2\%
(shaded areas) and the original potential (heavy line).
For this reconstruction, $\Omega_M = 0.3$ and
$V(\phi ) = V_0[1+\cos (\phi /f) ]$ (from Huterer \& Turner,
1998).
}
\label{fig:ht_recon}
\end{figure}

Both cosmologists and supernova theorists seem to use Eddington's
test to some degree.  It seems to me that the summary of the SN Ia
part of the meeting goes like this:  We don't know what SN Ia are;
we don't know how they work; but we believe SN Ia are very good
standardizeable candles.  I think what they mean is they have a
general framework for understanding a SN Ia, the thermonuclear
detonation of a Chandrasekhar mass white dwarf, and have failed
in their models to find a second (significant) parameter that
is consistent with the data at hand.  Cosmologists are persuaded
that the Universe is accelerating both because of the
SN Ia results and because this was the missing piece to a grander puzzle.

Not only have SN Ia led us to the acceleration of the Universe,
but also I believe they will play a major role in unraveling the
mystery of the dark energy.  The reason is simple:  we can be
confident that the dark energy was an insignificant component
in the past; it has just recently become important.  While, the
anisotropy of the CBR is indeed a cosmic Rosetta Stone, it is
most sensitive to physics around the time of decoupling.  (To
be very specific, the CBR power spectrum is almost identical
for all flat cosmological models with the same conformal age today.)
SNe Ia probe the Universe just around the time dark energy was becoming
dominant (redshifts of a few).  My student Dragan Huterer and I
(Huterer \& Turner, 1998)
have been so bold as to suggest that with 500 or so SN Ia with
redshifts between 0 and 1, one might be able to discriminate between
the different possibilities and even reconstruct the scalar potential
for the quintessence field (see Fig.~\ref{fig:ht_recon}).

\begin{acknowledgments}
My work is supported by the US Department of Energy and the
US NASA through grants at Chicago and Fermilab.
\end{acknowledgments}


\begin{thebibliography}{} 

\bibitem[]{} Bahcall, N. \& Fan, X. 1998, \apj, 504, 1.

\bibitem[]{} Bronstein, M.P. 1933, Phys. Zeit. der Sowjetunion, 3, 73.

\bibitem[]{} Bludman, S. \& Ruderman, M. 1977, \prl, 38, 255.

\bibitem[]{} Burles, S. \& Tytler, D. 1998a, \apj, 499, 699.

\bibitem[]{} Burles, S. \& Tytler, D. 1998b, \apj, 507, 732.

\bibitem[]{} Burles, S., Nollett, K., Truan, J. \& Turner, M.S. 1999,
\prl, in press.

\bibitem[]{} Caldwell, R., Dave, R., \& Steinhardt, P.J. 1998,
\prl, 80, 1582.

\bibitem[]{} Carlstrom, J. 1999, Physica Scripta, in press.

\bibitem[]{} Carroll, S. 1998, \prl, 81, 3067.

\bibitem[]{} Chaboyer, B. \etal\ 1998, \apj, 494, 96.

\bibitem[]{} Chiba, M. \& Yoshii, Y. 1998, \apj, in press (astro-ph/9808321).

\bibitem[]{} Coble, K., Dodelson, S. \& Frieman, J. A. 1996, \prd, 55, 1851.

\bibitem[]{} Dodelson, S., Gates, E.I. \& Turner, M.S. 1996, Science, 274, 69.

\bibitem[]{} Freese, K. \etal\ 1987, Nucl. Phys. B, 287, 797.

\bibitem[]{} Frieman, J., Hill, C., Stebbins, A. \& Waga, I. 1995, \prl, 75, 2077.

\bibitem[]{} Fukuda, Y. \etal\ (SuperKamiokande Collaboration) 1998, \prl,
81, 1562.

\bibitem[]{} Guerra, E.J., Daly, R.A. \& Wan, L. 1998, \apj, submitted
(astro-ph/9807249)

\bibitem[]{} Harvey, J. 1998, hep-th/9807213.

\bibitem[]{} Henry, P. 1998, in preparation.

\bibitem[]{} Huterer, D. \& Turner, M.S. 1998, \prl, submitted (astro-ph/9808133).

\bibitem[]{} Kamionkowski, M., Spergel, D.N. \& Sugiyama, N. 1994,
\apjl, 426, L57.

\bibitem[]{} Kochanek, C. 1996, \apj, 466, 638.

\bibitem[]{} Krauss, L. \& Turner, M.S., 1995, Gen. Rel. Grav., 27, 1137.

\bibitem[]{} Lineweaver, C. 1998, \apjl, 505, 69.

\bibitem[]{} Mohr, J., Mathiesen, B. \& Evrard, A.E. 1998, \apj, submitted.

\bibitem[]{} Mohr, J. \etal\ 1999, in preparation.

\bibitem[]{} Ostriker, J.P. \& Steinhardt, P.J. 1995, Nature 377, 600.

\bibitem[]{} Ozer, M. \& Taha, M.O. 1987, Nucl. Phys. B, 287, 776.

\bibitem[]{} Peebles, P.J.E. 1984, \apj, 284, 439.

\bibitem[]{} Perlmutter, S. \etal\ 1997, \apj, 483, 565.

\bibitem[]{} Perlmutter, S. \etal\ 1998, \apj, in press (astro-ph/9812133).

\bibitem[]{} Perlmutter, S., Turner, M.S. \& White, M. 1999,
\prl, submitted (astro-ph/9901052).

\bibitem[]{} Ratra, B. \& Peebles, P.J.E. 1988, \prd, 37, 3406.

\bibitem[]{} Riess, A. \etal\ 1998, \aj, 116, 1009.

\bibitem[]{} Sundrum, R. 1997, hep-th/9708329.

\bibitem[]{} Spergel, D. N. \& Pen, U.-L. 1997, \apjl, 491, L67.

\bibitem[]{} Turner, E.L. 1990, \apjl, 365, L43.

\bibitem[]{} Turner, M.S. 1983, \prd, 28, 1243.

\bibitem[]{} Turner, M.S. 1991, Physica Scripta, T36, 167.

\bibitem[]{} Turner, M.S. 1997, in Critical Dialogues in Cosmology,
ed. N. Turok (World Scientific, Singapore), p.~555.

\bibitem[]{} Turner, M.S. 1999, Physica Scripta, in press (astro-ph/9901109).

\bibitem[]{} Turner, M.S., Steigman, G. \& Krauss, L. 1984, \prl, 52, 2090.

\bibitem[]{} Turner, M.S. \& White, M. 1997, \prd, 56, R4439.

\bibitem[]{} Turner, M.S. \& Wilczek, F. 1982, Nature, 298, 633.

\bibitem[]{} Weinberg, S. 1989, Rev. Mod. Phys., 61, 1.

\bibitem[]{} Vilenkin, A. 1984, \prl, 53, 1016.

\bibitem[]{} Waga, I. \& Miceli, A.P.M.R. 1998, astro-ph/9811460.

\bibitem[]{} White, S.D.M. \etal\ 1993, Nature 366, 429.

\bibitem[]{} Zel'dovich, Ya.B. 1967, JETP, 6, 316.


\end{thebibliography}
\end{document}